\documentclass[aps,showpacs,twocolumn]{revtex4}
\usepackage{mathrsfs}
\usepackage{amssymb}
\usepackage{amsmath}
\usepackage{bm}
\usepackage{graphics}
\usepackage{natbib}

\begin{document}
\title{Well-defined Fano effect in the Andreev reflection process of a parallel double-quantum-dot structure}
\author{Xiao-Qi Wang$^1$}
\author{Shu-Feng Zhang$^2$}
\author{Yu Han$^3$}

\author{Wei-Jiang Gong$^1$}\email[Author to whom correspondence should be
addressed. Email address: ]{gwj@mail.neu.edu.cn}

\affiliation{1. College of Sciences, Northeastern University, Shenyang 110819, China\\
2. School of Physics and Technology, University of Jinan, Jinan, Shandong 250022, China \\
3. School of Physics, Liaoning University, Shenyang 110036, China}
\date{\today}

\begin{abstract}
We investigate the Andreev reflection in the parallel double-quantum-dot structure, by considering one metallic lead and one $s$-wave superconductor to couple to the quantum dots simultaneously. It is found that the Fano lineshhape has opportunities to appear in the linear conductance spectrum of the Andreev reflection, which can be reversed by tuning the dot level or local magnetic flux. However, the property of the Fano effect is very complicated, in comparison with the normal electron tunnelling case. This is manifested as the special Fano form of the linear-conductance expression and the interference manner among the Feynman paths. We believe that this work can be helpful for understanding the Fano interference in the Andreev reflection process.
\end{abstract}
\pacs{74.45.+c, 73.63.Kv, 73.23.-b} \maketitle

\bigskip

\section{Introduction}

The Fano effect, well-known for the asymmetric lineshapes in the spectra concerned, is an important physics phenomenon in many fields, such as optics, atom and molecule physics, and condensed matter physics. Its underlying reason is the interference between resonant and reference processes\cite{refFano}. In low-dimensional semiconductor systems, electronic transport is governed by quantum coherence. When the resonant and nonresonant tunneling channels are achieved, the Fano lineshapes have opportunities to appear in the transport spectra. One typical system for observing the Fano effect is the coupled-QD structure, which provides
multiple channels for electronic coherent transmission. In appropriate parameter region, one or a few channels serve as the
resonant paths for electron tunneling and the others are the nonresonant ones, leading to the occurrence of the Fano interference\cite{RMP}.
\par
Recently, Fano effect manifests itself in QD structures by the experimental observation of the asymmetric lineshape in the
conductance spectrum\cite{refGores,refGordon,refIye,refSato,refKobayashi}. Meanwhile, the relevant theoretical investigations involve various QD structures,
for example, one or two QDs embedded in an Aharonov-Bohm ring\cite{refBulka,refKonig1,refKonig2,refGefen,refGefen2,refKash,refABring},
double QDs in different coupling manners\cite{refKang,refBai,refOrellana,refZhu,refDing}. According
to these theoretical results, the Fano effect in QD structures exhibits some peculiar behavoirs in electronic transport process, in
contrast to the conventional Fano effect. These include the tunable Fano lineshape by the magnetic or electrostatic fields applied on
the QDs\cite{refKang,refBai,refOrellana,refZhu}, the Kondo resonance associated Fano effect\cite{refBulka,refKonig1,refDing},
Coulomb-modification on the Fano effect\cite{refZhu2}, the spin-dependent Fano effect\cite{spin}, and the relation between the dephasing time and the Fano parameter
$q$\cite{refClerk}. In addition, the improvement of the thermoelectric efficiency induced by the Fano effect has been demonstrated\cite{Thermo}.

\par
At the same time, the developments in the microfabrication technology encourage scientists to take their interest in the mesoscopic systems with the coupling between normal metal and superconductors (SCs) in the past decades\cite{Andreev1,Andreev2,Andreev3,Andreev4,Andreev5,Andreev5(1),Andreev5(2)}. It is known that in such systems, the so-called Andreev reflection (AR) is allowed to take place at the metal-SC interface due to the appearance
of a new energy scale, i.e., the superconducting pairing energy $\Delta$\cite{Andreev6}. This phenomenon can be described as follows. If an incoming electron
from the normal metal incidents to the SC, it will be reflected as a hole, thereby transferring a Cooper pair into the SC. When the coupled QDs are embedded between the middle of the normal metal and SC, various intricate phenomena have been observed, such as the appearance of the Shiba states. Moreover,
since the SC is a the natural source of the entanglement electrons, the
multi-terminal hybrid systems have been proposed to be promising candidates for Cooper pair splitting, by manipulating the crossed AR between the normal metals\cite{Andreev7,Andreev8}.

\par
Following the developments in the above two aspects, the Fano effect in the AR process has attracted extensive attention\cite{Fano-A1,Fano-A2,Wangbg,Liyx,Molecule,Two,Poland}. As a result, some results have been observed. Peng $et$ $al.$ discussed the AR in a four-QD Aharonov-Bohm interferometer, and found that in this system, the reciprocity relation for conductance $\mathcal{G}(\varphi)=\mathcal{G}(-\varphi)$ holds. Moreover, the antiresonance point appears in the AR conductance spectrum with the adjustment of the systematic parameters\cite{Wangbg}. For the parallel-coupled double-QDs with spin-flip scattering, the Fano resonance has also been observed though it is relatively weak. And the spin-flip scattering enables to induces the splitting of the Fano peak\cite{Liyx}. Some other groups reported the AR in a normal metal-molecule-SC junction using a first-principles
approach. They demonstrated that the presence of the side group in the molecule can lead to a Fano resonance in the AR\cite{Molecule}. In addition, it has been showed that when the T-shaped double QDs is inserted in the metal-SC junction, two separate Fano structures appear in the gate-voltage dependence of the AR
process\cite{Two}. Decoherence takes effect to the Fano lineshapes in the T-shaped double QDs coupled between the metal and
SC, when an additional terminal is introduced\cite{Poland}. These results indeed suggest that the Fano effect in the AR is very interesting. However, to clarify its more physics picture, new discussion is still necessary.
\par
In the present work, we investigate the Fano interference in the AR process of the parallel double-QD structure, by considering one metallic lead and one SC to couple to the QDs respectively. It is found that Fano lineshhape has opportunities to appear in the linear AR conductance spectrum, which can be reversed by tuning the QD level and local magnetic flux. However, the condition of the Fano resonance is completely different from that in the normal electron transport case.

\par
The rest of the paper is organized as follows. In Sec. \ref{theory}, the model Hamiltonian to describe the
electron motion in the parallel double-QD structure is introduced
firstly. Then a formula for the linear AR conductance is derived by
means of the nonequilibrium Green function technique. In
Sec. \ref{result2}, the calculated results about the linear
conductance spectrum are shown. Then a discussion focusing on the
formation of Fano lineshape is given. Finally, the main results are
summarized in Sec. \ref{summary}.
\begin{figure}
\centering \scalebox{0.43}{\includegraphics{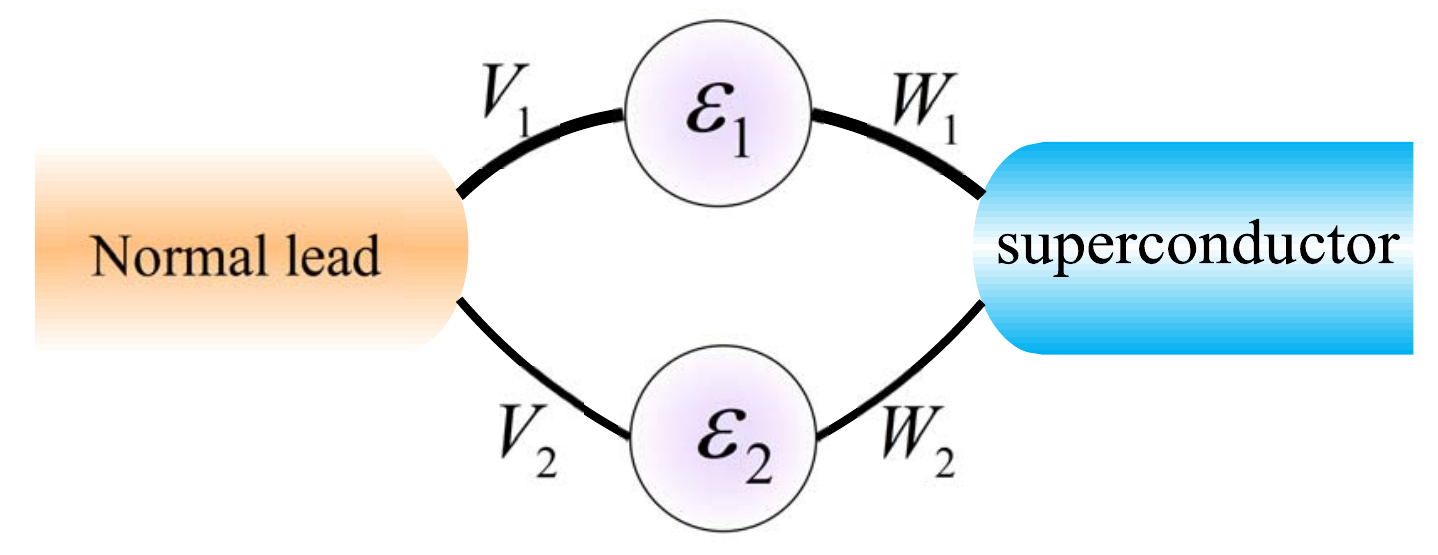}} \caption{
Schematic of one parallel double-QD structure. In this geometry, the QDs are coupled to one normal metallic lead and one $s$-wave SC simultaneously.     \label{structure0}}
\end{figure}
\section{model\label{theory}}
The parallel double-QD structure that we consider is illustrated in Fig.
\ref{structure0}. The Hamiltonian to
describe the electronic motion in this double-QD geometry
reads
\begin{equation}
H=H_{n}+H_s+H_{d}+H_{nT}+H_{sT}.   \label{1}
\end{equation}
The first two terms (i.e., $H_n$ and $H_s$) are the Hamiltonians for the electrons
in the normal metallic lead and SC, respectively:
\begin{eqnarray}
H_{n}&&=\underset{k\sigma}{\sum }\varepsilon
_{k}c_{k\sigma}^\dag c_{k \sigma},\notag\\
H_{s}&&=\underset{k\sigma}{\sum }\epsilon
_{k}a_{k\sigma}^\dag a_{k\sigma}+\sum_k(\Delta_ka^\dag_{k\uparrow}a^\dag_{-k\downarrow}+\mathrm{h.c.}).\label{2}
\end{eqnarray}
$c_{k\sigma}^\dag$ $( c_{k\sigma})$ is an operator to create
(annihilate) an electron of the continuous state $|k\sigma\rangle$ in the normal metallic lead,
and $\varepsilon _{k}$ is the corresponding single-particle
energy. $a_{k\sigma}^\dag$ $(a_{k\sigma})$ is the creation (annihilation) operator in the SC.
$\epsilon_{k}$ is the corresponding energy, and $\Delta_k$ is the superconducting pairing potential.

The third term describes the electron in the double QDs. It
takes a form as
\begin{equation}
H_{d}=\sum_{j=1}^{2}\varepsilon_{j}d_{j\sigma}^\dag d_{j\sigma},\label{3}
\end{equation}
where $d^{\dag}_{j\sigma}$ $(d_{j\sigma})$ is the creation (annihilation)
operator of electron in the $j$th QD, and $\varepsilon_j$ denotes the
electron level in the corresponding QD. We here assume that only one
level is relevant in each QD, since we mainly take our interest in the AR governed by the quantum coherence. The last two terms in the Hamiltonian
describe the QD-lead and QD-SC couplings. They are
given by
\begin{eqnarray}
H_{nT}&&=\underset{k\sigma,j }{\sum } V_{jk}c_{k\sigma}^\dag
d_{j\sigma}+{\mathrm {h.c.}} , \notag\\
H_{sT}&&=\underset{k\sigma,j }{\sum } W_{jk}a_{k\sigma}^\dag
d_{j\sigma}+{\mathrm {h.c.}} .
\end{eqnarray}
$V_{jk}$ ($W_{jk}$) denotes the QD-lead (QD-SC) coupling strength with
$j=1,2$. In the presence of local magnetic flux through this geometry, phase shift will attached to the QD-lead coupling coefficients, which gives that
$V_{1k}=|V_{1k}|e^{i\varphi/2}$ and
$V_{2k}=|V_{2k}|e^{-i\varphi/2}$. The phase shift $\Phi$ is associated
with the magnetic flux $\varphi$ threading the system by a relation
$\varphi=2\pi\Phi/\Phi_{0}$, in which $\Phi_{0}=h/e$ is the flux
quantum\cite{refIye,Mag2}.
\par
In this work, we are interested in the AR process in the superconducting gap, hence the approximation of $\Delta_{k}\to\infty$ is feasible\cite{largegap,largegap1}. It should be emphasized that in such a case,
as the Bogoliubov quasiparticle in the SC is inaccessible,
tracing out of the degree of freedom of the SC does not induce any dissipative dynamics in the QDs system and it can be performed exactly. The resulting
Hamiltonian dynamics of the considered structure can be redescribed by the
following effective Hamiltonian:
$H=H_n+\tilde{H}_d+H_{nT}$, where
\begin{small}
\begin{eqnarray}
\tilde{H}_d&=&\sum_{\sigma,j=1}^{2}\varepsilon _{j}d_{j\sigma}^\dag d_{j\sigma}-\sum_{j=1}^{2}\Delta_j d^\dag_{j\uparrow}d^\dag_{j\downarrow}
+\sqrt{\Delta_1\Delta_2}d^\dag_{1\uparrow}d^\dag_{2\downarrow}\notag\\
&&+\mathrm{h.c.}.
\end{eqnarray}
\end{small}
$\Delta_{j}$, defined by $\Delta_{j}=\pi \sum_k |W_{jk}|^2 \delta(\omega-\epsilon_k)$, represents the effective superconducting pairing potential in QD-$j$ induced by the proximity effect between the SC and it.
\par
We now proceed to calculate the current passing through the metallic
lead, which can be defined as $J=-e\langle \dot{\hat{N}}\rangle$ with $\hat{N}=\sum_{k\sigma}c_{k\sigma}^\dag c_{k \sigma}$. Using the Heisenberg equation of
motion, the current can be rewritten as $J=-e\sum_{j,k\sigma}[V_{jk}G^<_{jk,\sigma}(t,t)+c.c]$, where $G^<_{jk,\sigma}(t,t')=i\langle c^\dag_{k\sigma}(t')d_{j\sigma}(t)\rangle$ is the lesser Green's function. With the help of the
Langreth continuation theorem and taking the Fourier
transformation, we have\cite{Formula} 
\begin{equation}
J={e\over h}\int dE \mathrm{Tr}\{\sigma_3\Gamma[(G^r-G^a)f(E)+G^<]\}
\end{equation}
in which $f(E)$ is the Fermi distribution function. $G^{r,a,<}$ are the retarded, advanced, and lesser Green's
functions in the Nambu representation, which are defined as $G^r(t,t')=-i\theta(t-t')\langle\{\Psi(t),\Psi^\dag(t')\}\rangle$
and $G^<(t,t')=i\langle\{\Psi^\dag(t')\Psi(t)\}\rangle$ with $G^a=[G^r]^\dag$.
The field operator involved, i.e., $\Psi$, is given by  $\Psi=[d_{1\uparrow},d^\dag_{1\downarrow},d_{2\uparrow},d^\dag_{2\downarrow}]^T$.
$\Gamma$ is the linewidth matrix function of
the metallic lead, which describes the coupling strength between the lead and the QDs. If the lead is manufactured by two-dimensional electron gas, the elements of $\Gamma$ will be independent of energy.
\par
It is not difficult to find that for calculating the current, one must obtain
the expressions of the retarded and lesser Green's functions.
The retarded Green's function can be in principle obtained
from the Dyson's equation. Via a straightforward derivation, the matrix of the retarded Green's function can be written out, i.e.,
\begin{small}
\begin{eqnarray}
&&[G^r(E)]^{-1}=\notag\\
&&\left[\begin{array}{cccc} g_{1e}(E)^{-1} &\Delta_1 & {i\over2}\Gamma_{12,e}&-\sqrt{\Delta_1\Delta_2}\\
\Delta_1& g_{1h}(E)^{-1}&-\sqrt{\Delta_1\Delta_2}&{i\over2}\Gamma_{12,h}\\
{i\over2}\Gamma_{21,e}&-\sqrt{\Delta_1\Delta_2}&g_{2e}(E)^{-1}& \Delta_2\\
-\sqrt{\Delta_1\Delta_2}&{i\over2}\Gamma_{21,h}&\Delta_2& g_{2h}(E)^{-1}
  \end{array}\right]\ \label{green},
\end{eqnarray}
\end{small}
where $g_{je(h)}=[E\mp\varepsilon_j+{i\over2}\Gamma_{jj,e}]^{-1}$. $\Gamma_{jl,e}$ and $\Gamma_{jl,h}$ are defined as $\Gamma_{jl,e}=2\pi\sum_k V^*_{jk}V_{lk}\delta(E-\varepsilon_k)$ and $\Gamma_{jl,h}=2\pi\sum_k V_{jk}V^*_{lk}\delta(E+\varepsilon_k)$, respectively. Within the wide-band approximation of the lead, we will have $\Gamma_{jj,e}=\Gamma_{jj,h}$. And then, the matrixes of $\Gamma_e$ and $\Gamma_h$ can be respectively expressed as
\begin{small}
\begin{eqnarray}
\Gamma_{e}=\left[\begin{array}{cccc} \Gamma_{1} &0& \sqrt{\Gamma_{1}\Gamma_2}e^{-i\varphi}&0\\
0& 0&0&0\\
\sqrt{\Gamma_{1}\Gamma_2}e^{i\varphi}&0&\Gamma_{2}& 0\\
0&0&0& 0
  \end{array}\right]\ \label{green}
\end{eqnarray}
\end{small}
and
\begin{small}
\begin{eqnarray}
\Gamma_{h}=\left[\begin{array}{cccc} 0&0&0&0\\
0&\Gamma_{1}&0&\sqrt{\Gamma_{1}\Gamma_2}e^{i\varphi}\\
0&0&0& 0\\
0&\sqrt{\Gamma_{1}\Gamma_2}e^{-i\varphi}&0&\Gamma_{2}
  \end{array}\right]\ \label{green}.
\end{eqnarray}
\end{small}
As for the lesser Green's function, it can be deduced by using Keldysh equation
$G^<=G^r\Sigma^<G^a$, where
\begin{eqnarray}
\Sigma^<=\left[\begin{array}{cc} \Sigma^<_{11} &\Sigma^<_{12}\\
\Sigma^<_{21}& \Sigma^<_{22}
  \end{array}\right]\
\end{eqnarray}
with
$
\Sigma^<_{jl}=\left[\begin{array}{cc} i\Gamma_{jl,e}f(E-eV) &0\\
0& i\Gamma_{jl,h}f(E+eV)
  \end{array}\right].\
$
After the derivation above, the electronic current
can be further simplified, i.e.,
\begin{small}
\begin{eqnarray}
J={2e\over h}\int dE\{T_A(E)[f(E-eV)-f(E+eV)]\}.
\end{eqnarray}
\end{small}
where $T_A=\mathrm
{Tr}[\Gamma_eG^r(E)\Gamma_hG^a(E)]$ is the AR function.
In the case of zero temperature limit, the current formula can be reexpressed, yielding
$J={2e\over h}\int _{-eV}^{eV}T_A(E)dE$. It is evident that $T_A(E)$ is an important quantity to evaluate the electronic current driven by the AR. After a complicated derivation, we can write out the analytical its expression, i.e., $T_A(E)=|\tau(E)|^2$ with the AR coefficient being
\begin{widetext}
\begin{equation}
\tau(E)=\frac{E^2(e^{-i\varphi}\Gamma_2\Delta_2+e^{i\varphi}\Gamma_1\Delta_1-\sqrt{\Gamma_1\Gamma_2\Delta_1\Delta_2})-(e^{-i\frac{\varphi}{2}}\varepsilon_1\sqrt{\Gamma_2\Delta_2}-e^{i\frac{\varphi}{2}}\varepsilon_2\sqrt{\Gamma_1\Delta_1})^2}{\det|[G^r(E)]^{-1}|}.
\end{equation}
\end{widetext}
At the zero-bias limit, we will have $J=\mathcal{G}\cdot V$ with the linear conductance given by
\begin{equation}
\mathcal{G}={4e^2\over h}T_A|_{E=0}.\label{linear}
\end{equation}
Note that at equilibrium, the chemical potential in the normal metallic lead has been assumed to be zero. As a result, the expression of $\tau$ is written as
\begin{small}
\begin{eqnarray}
\tau|_{E=0}=\frac{-4(e^{-i\frac{\varphi}{2}}\varepsilon_1\sqrt{\Gamma_2\Delta_2}-e^{i\frac{\varphi}{2}}\varepsilon_2\sqrt{\Gamma_1\Delta_1})^2}{ (\varepsilon_1\Gamma_2+\varepsilon_2\Gamma_1)^2
+4(\varepsilon_1\Delta_2+\varepsilon_2\Delta_1)^2+4\varepsilon_1^2\varepsilon_2^2+{\cal C}} \notag
\end{eqnarray}
\end{small}
where ${\cal C}=|\sqrt{\Gamma_1\Delta_2}e^{i\varphi/2}+\sqrt{\Gamma_2\Delta_1}e^{-i\varphi/2}|^4$. At the case of $\Delta_j=\Delta$, there will be
\begin{small}
\begin{eqnarray}
\tau|_{E=0}=\frac{-4\Delta(e^{-i\frac{\varphi}{2}}\varepsilon_1\sqrt{\Gamma_2}-e^{i\frac{\varphi}{2}}\varepsilon_2\sqrt{\Gamma_1})^2}{ (\varepsilon_1\Gamma_2+\varepsilon_2\Gamma_1)^2
+4\Delta^2(\varepsilon_1+\varepsilon_2)^2+4\varepsilon_1^2\varepsilon_2^2+{\cal C}} \notag
\end{eqnarray}
\end{small}
in which ${\cal C}=\Delta^2|\sqrt{\Gamma_1}e^{i\varphi/2}+\sqrt{\Gamma_2}e^{-i\varphi/2}|^4$.

\begin{figure}
\centering \scalebox{0.43}{\includegraphics{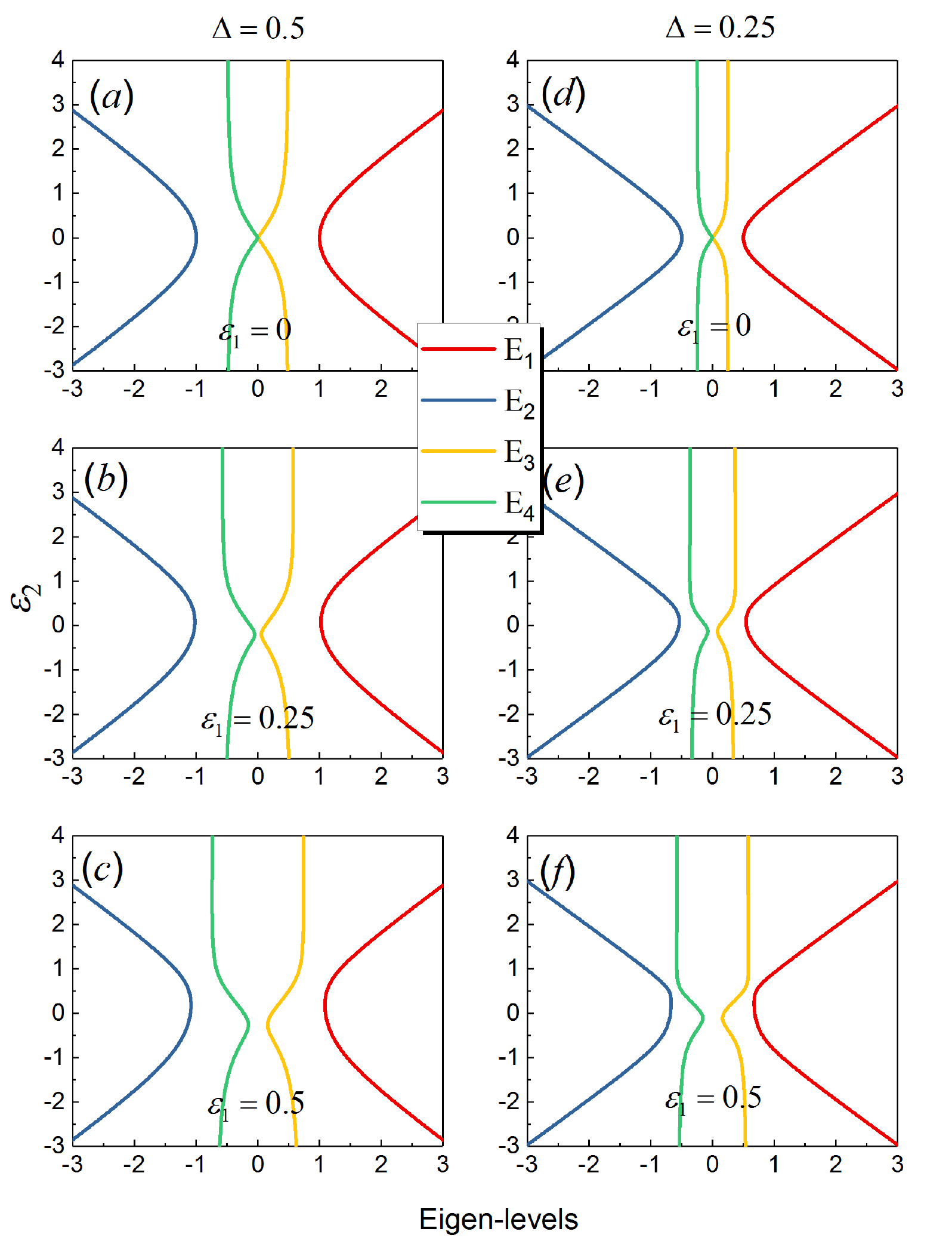}} \caption{
Eigen-levels of the double QDs in the Nambu representation, due to their coupling via one SC. \label{fanoline1}}
\end{figure}


\begin{figure}
\centering \scalebox{0.16}{\includegraphics{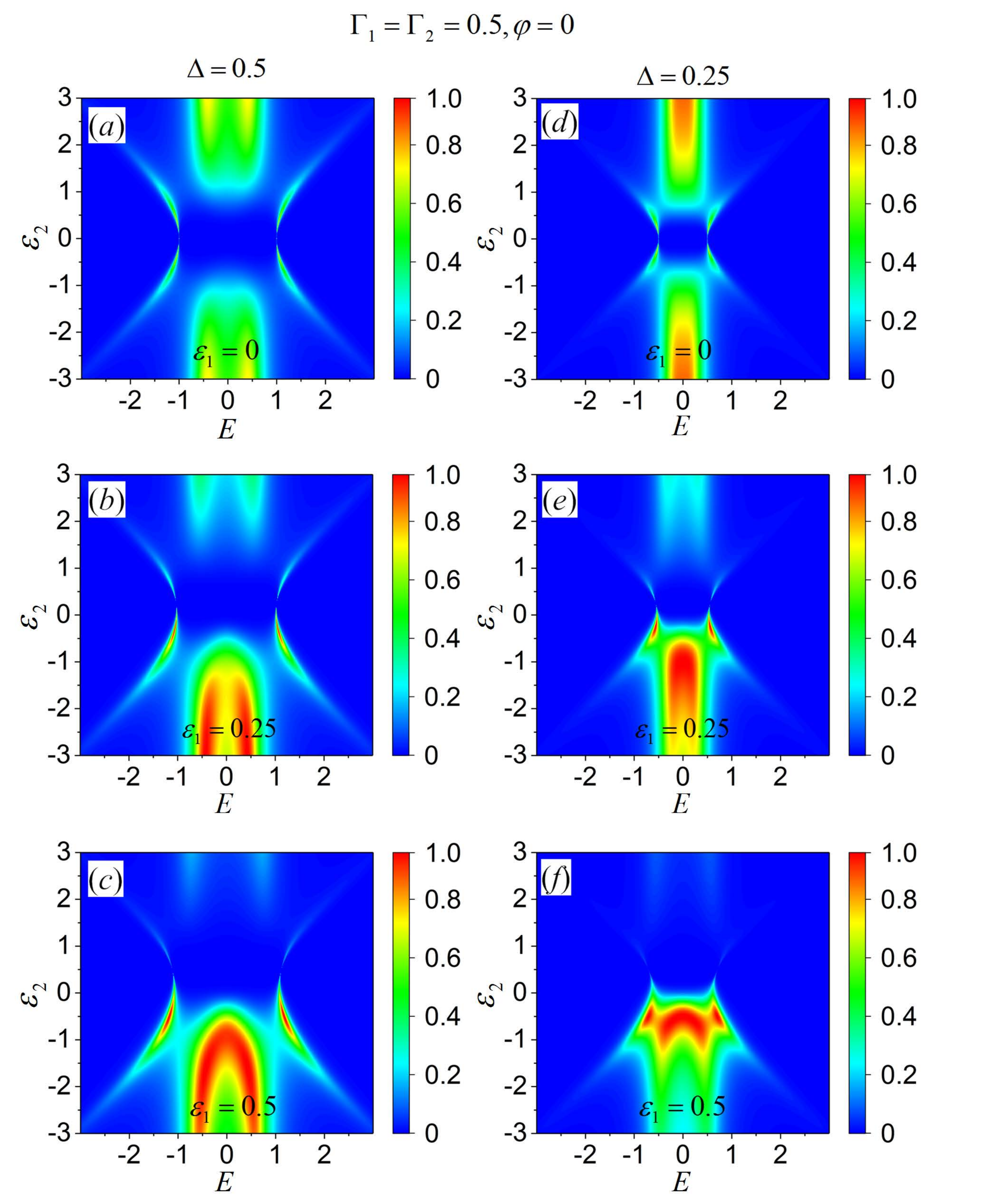}} \caption{
Spectra of the AR in the parallel double-QD structure. The level of QD-1 is taken to be $\varepsilon_1=0$, $0.25$, and $0.5$. The identical QD-lead couplings is fixed with $\Gamma=0.5$. (a)-(c) Results in the cases of $\Delta=0.5$, and (d)-(f) correspond to the results of $\Delta=0.25$. \label{fanoline}}
\end{figure}

\section{Numerical results and discussions \label{result2}}
Following the formulation developed in the previous section, we continue to
perform the numerical calculation to investigate the AR spectrum of the parallel double-QD structure with one SC.
\par
Before our discussion, we try to solve the eigenlevels of the double QDs in the Nambu representation, which play nontrivial roles in governing the AR properties of this system. After calculation, we obtain the four eigenlevels as $E_1=-E_4={-1\over \sqrt{2}}\sqrt{\varepsilon_1^2+\varepsilon_2^2+4\Delta^2+\Omega}$ and $E_2=-E_3={-1\over \sqrt{2}}\sqrt{\varepsilon_1^2+\varepsilon_2^2+4\Delta^2-\Omega}$ with $\Omega=\sqrt{(\varepsilon_1^2-\varepsilon_2^2)^2+4\Delta^2(\varepsilon_1-\varepsilon_2)^2+16\Delta^4}$.
With the help of this result, we can find the leading feature of the eigenlevels. To be concrete, in the case of $\varepsilon_1=\varepsilon_2=0$, the level degeneracy has an opportunity to take place between $E_2$ and $E_3$, where $E_1$ and $E_4$ are away from each other by $2\Delta$. Detailed information of the eigenlevels can be observed in Fig.2. It shows that $E_{1(2)}$ (or $E_{3(4)}$) change in opposite manners with the adjustment of $\varepsilon_2$. When $\varepsilon_2$ departs from its zero value, the absolute value of $E_1$ ($E_4$) increases linearly, but $E_2$ ($E_3$) tends to keep changeless. This change manner is reversed when $\varepsilon_2$ varies around its zero value. In addition, one can notice that reducing $\Delta$ mainly leads to the decrease of the minimum of $E_1$ and the maximum of $E_2$.

Next, we proceed to investigate the AR behaviors in our structure by considering two coupling manners, i.e., the case of symmetric QD-lead coupling where $\Gamma_j=\Gamma$ and $\Delta_j=\Delta$ and the case of asymmetric QD-lead coupling where $\Gamma_1=10\Gamma_2$ and $\Delta_1=10\Delta_2$.
\subsection{Case of symmetric QD-lead couplings}
\par
We first consider the case of symmetric QD-lead couplings and present the AR spectra of the parallel double-QD structure. In this case, the AR coefficient can be directly written as
\begin{small}
\begin{equation}
\tau(E)=\frac{(2\cos\varphi-1)E^2-(e^{-i\varphi/2}\varepsilon_1-e^{i\varphi/2}\varepsilon_2)^2}{\det|[G^r(E)]^{-1}|}\Gamma\Delta. \label{ddd}
\end{equation}
\end{small}
\par
In the absence of magnetic flux, there will be
$\tau(E)=\frac{1}{\det|[G^r(E)]^{-1}|}[E^2-(\varepsilon_1-\varepsilon_2)^2]\Gamma\Delta$.
This means that the antiresonance points will appear in the AR spectrum in the situation of $E=\pm(\varepsilon_1-\varepsilon_2)$.
The AR numerical results are shown in Fig.3, where $\Gamma=0.5$. With respect to the effective pairing potentials in the QDs, they are assumed to be $\Delta=0.5$ and $0.25$, respectively. Fig.3(a)-(c) show the results of $\varepsilon_1=0$, $0.25$, and $0.5$, in the case of $\Delta=0.5$. In Fig.3(a) we see that the eigenlevels of the double QDs provide channels for the AR process. This is manifested as that the peaks of the AR spectra are consistent with the eigenlevels of the double QDs. However, the value of $T_A(E)$ cannot reach unity in the case of $E=E_j$. It also shows that $E_1$ and $E_4$ make little contribution to the AR. And the AR peaks related to $E_2$ and $E_3$ disappear in the region of $-1.0<\varepsilon_2<1.0$. In Fig.3(b)-(c), one can find that when the level of QD-1 shifts away from the zero energy point, the asymmetry of the AR spectrum becomes more apparent with the change of $\varepsilon_2$. In the case of $\varepsilon_1=0.5$, the AR is almost suppressed in the region of $\varepsilon_2>0$. Instead, in the region of $\varepsilon_2<0$, the AR is enhanced to much degree. and the AR peaks contributed by $E_2$ and $E_3$ get crossed. As a result, the AR becomes resonant at the point of $E=0$.
Next, when the superconducting pairing potential in the QD decreases with $\Delta=0.25$, the AR spectra are narrowed seriously and the AR phenomenon mainly occurs in the region of $|E|<0.5$. Besides, the contribution of $E_1$ and $E_4$ becomes more weak, as shown in Fig.3(d)-(f). The other result lies in that the
relation between the double-QD eigenlevels and the AR peaks becomes relatively ambiguous. All these results indicate that compared with the single-electron tunneling, the AR peaks relate to the eigenlevels of the QDs in the Nambu representation in an alternative way. Meanwhile, the co-existence of two antiresonance conditions makes nontrivial impact on the AR spectrum of this system.
\begin{figure}[htb]
\centering \scalebox{0.16}{\includegraphics{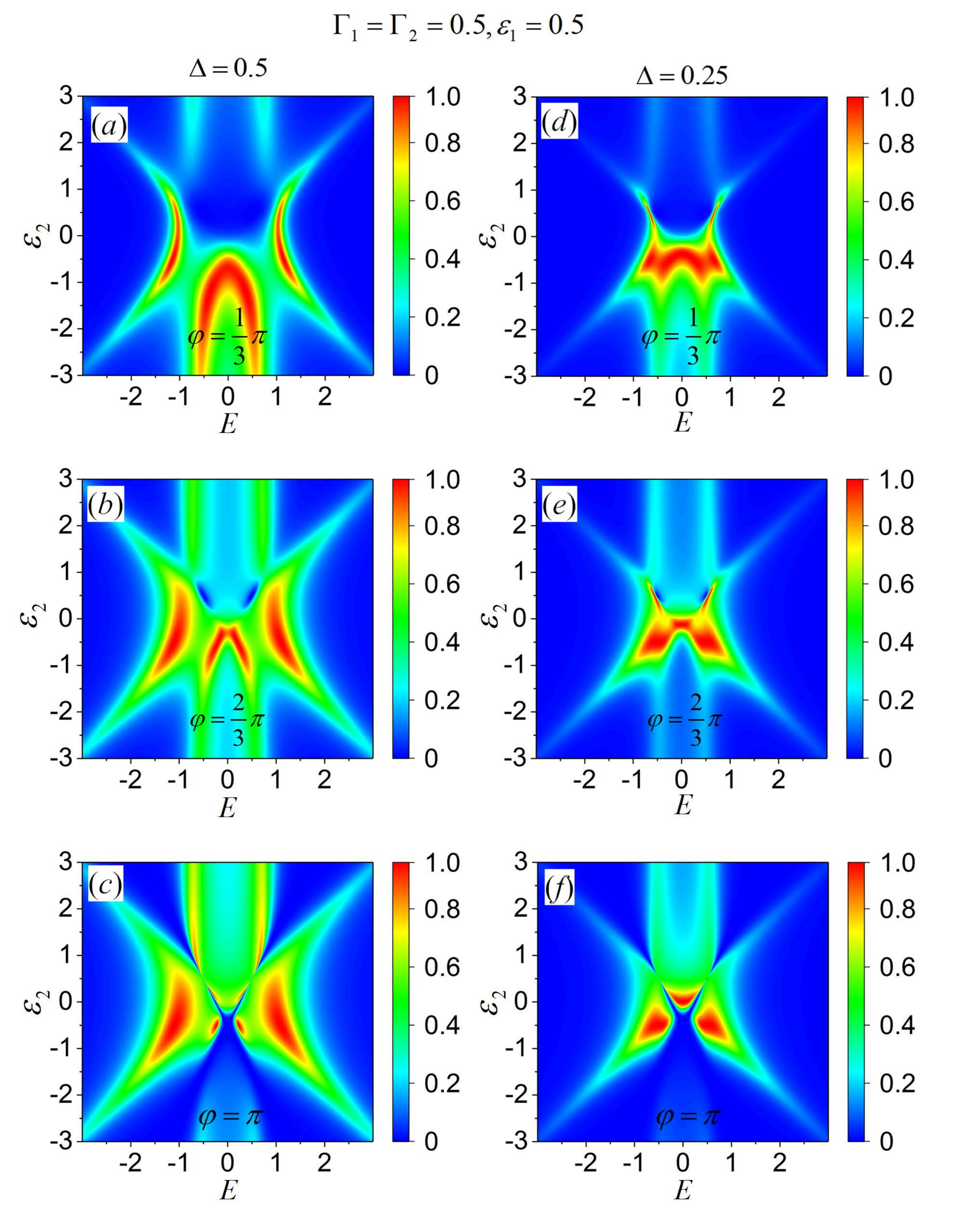}} \caption{ AR spectra affected by the local magnetic flux through this system. The level of QD-1 is fixed with $\varepsilon_1=0.5$. Other structural parameters are the same as those in Fig.2, respectively. \label{fanoline1}}
\end{figure}
\par
In Fig.4, we introduce local magnetic flux through this ring junction to investigate the change of AR properties. From Eq.(\ref{ddd}), it can be observed that in the case of $\varphi=\pi$, $\tau(E)$ has an opportunity to be equal to zero, corresponding to the occurrence of antiresonance phenomenon at the positions of $E=\frac{\pm 1}{\sqrt{3}}(\varepsilon_1+\varepsilon_2)$. Fig.4(a)-(c) show the results of $\varphi={\pi\over3}$, $2\pi\over3$, and $\pi$, respectively, in the case of $\varepsilon_1=0.5$ with $\Delta=0.5$. And the result of $\Delta=0.25$ are exhibited in Fig.4(d)-(f). One can readily find that with the increase of magnetic flux, the AR spectrum experiences obvious changes. Firstly, the $E_1$-contributed AR peaks are strengthened and widened apparently. On the other hand, the magnetic flux modifies the role of $\varepsilon_2$. As shown in Fig.4(c), when $\varepsilon_2$ tunes in the positive-energy region, the AR begins to be more active. Also, due to the occurrence of the antiresonances, the edges of the respective parts of the AR spectrum can be clearly seen in the case of $\varphi=\pi$. Next for the case of $\Delta=0.25$, the leading properties of the AR spectra are basically similar to those in the above case [See Fig.4(d)-(f)]. The difference lies in the narrowness of the AR spectra. This inevitably re-regulates the AR peaks. As shown in Fig.4(f), there are three peaks survived in the AR spectrum in the case of $\varphi=\pi$.

\begin{figure}
\centering \scalebox{0.42}{\includegraphics{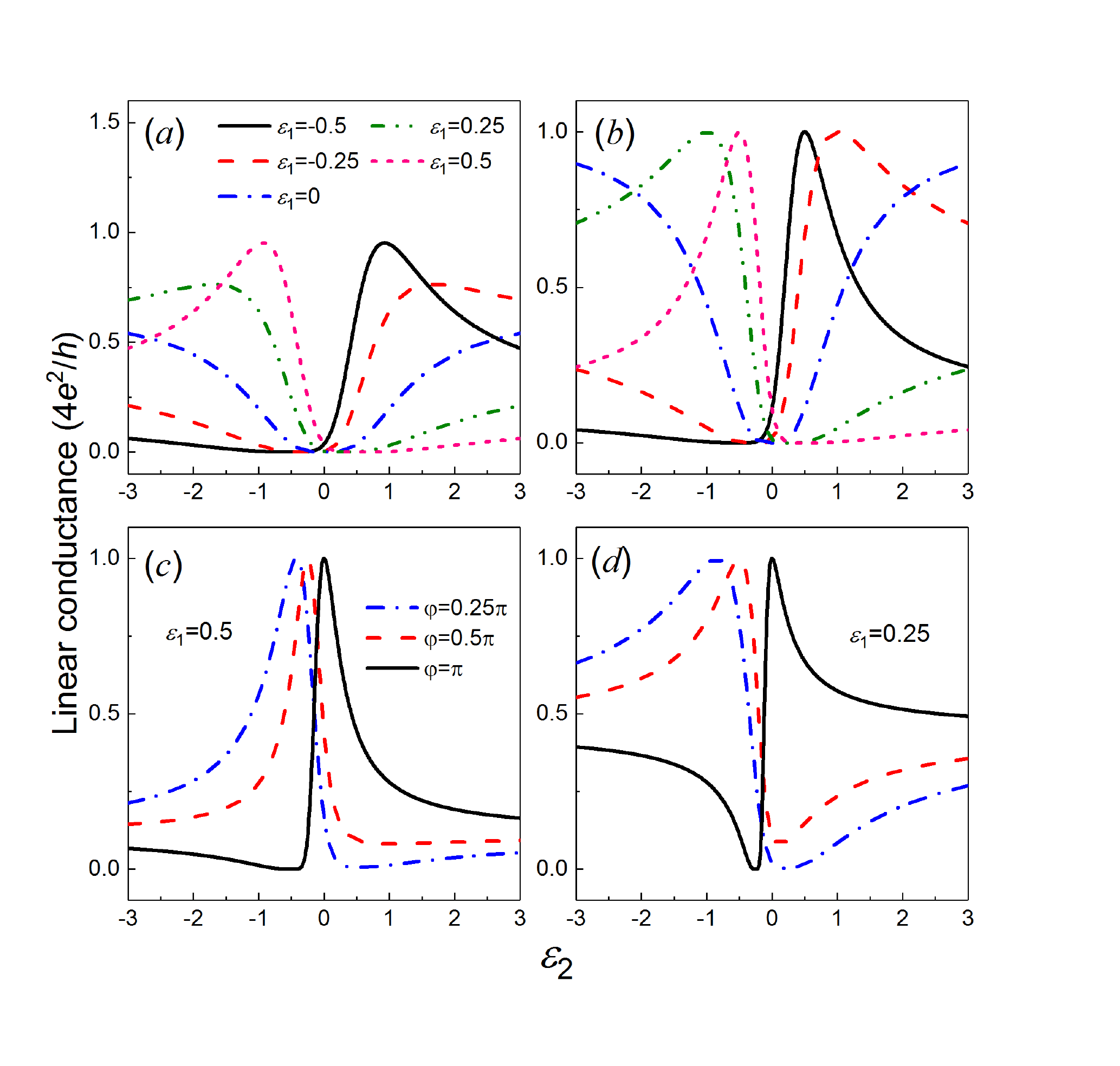}} \caption{
Linear-conductance spectra of the AR process in the parallel double-QD structure. The QD-lead coupling strengths are $\Gamma=0.5$. (a)-(b) Results in the cases of $\Delta=0.5$ and $0.25$, respectively. The level of QD-1 is taken to be $\varepsilon_1=\pm0.5$, $\pm0.25$, and $0$. (c)-(d) Influence of magnetic flux on the conductance spectra in the case of $\Delta=0.25$. \label{fanoline}}
\end{figure}

\par
In view of the above AR results, we would like to concentrate on its linear conductance spectrum to clarify the AR property in this structure. According to Eq.(\ref{linear}), the linear conductance is related to the AR ability in the case of $E=0$. In Fig.5, we plot the linear conductance curves, by taking $\Delta=0.5$ and $0.25$, respectively. Firstly, Fig.5(a) show the results of the fixed level of QD-1, where $\varepsilon_1=\pm0.5$, $\pm0.25$, and $0$. We can clearly find that in the case of $\varepsilon_1=0.5$, one Fano lineshape appears in the linear conductance spectrum, with the Fano peak and antiresonance at the positions of $\varepsilon_d=0$ and $\varepsilon_d=1.0$. When the level of QD-1 shifts to $\varepsilon_1=-0.5$, the Fano lineshape is reversed. This exactly means that in the linear AR process of the parallel double-QD structure, Fano interference can also take place. And its mode can be tuned with the help of the electric method. Next, if we choose $\varepsilon_1=\pm0.25$, there also exist the Fano lineshapes in the conductance spectra. However, no Fano resonance peak survives in such cases. If $\varepsilon_1=0$, only one symmetric lineshape can be observed, though the occurrence of the antiresonance at the position of $\varepsilon_d=0.5$. Next, if $\Delta$ decreases to 0.25, one can from Fig.5(b) find that the Fano resonance is allowed to take place as well, independent of the different $\varepsilon_1$. Therefore, the effective pairing potentials in the QDs play their special roles in driving the Fano resonance in the linear AR process. In addition, as shown in Fig.5(c)-(d), the local magnetic flux can effectively modify the Fano lineshape in the conductance spectrum. Until $\varphi=\pi$, the Fano lineshape is reversed. Note, however, that differently from the electric tuning, the Fano peak in the case of $\varphi=\pi$ becomes narrow after the reversal.

\par
\subsection{Case of asymmetric couplings}
\begin{figure}[htb]
\centering \scalebox{0.16}{\includegraphics{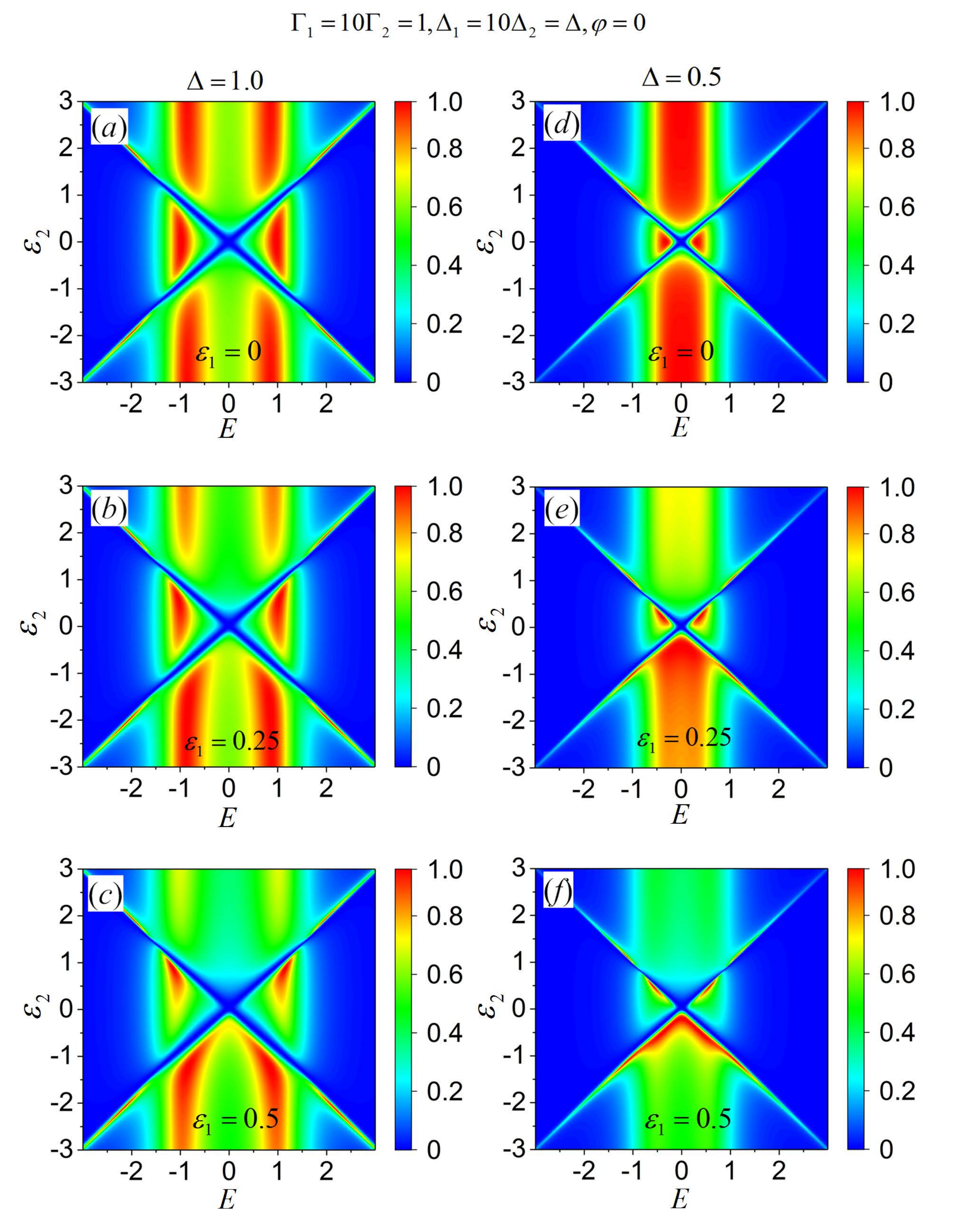}} \caption{
AR spectra in the parallel double-QD structure. The level of QD-1 is taken to be $\varepsilon_1=0$, $0.25$, and $0.5$. The QD-lead coupling strength is $\Gamma_{1}=10\Gamma_2=1.0$. (a)-(c) Results in the cases of $\Delta_1=1.0$, and (d)-(f) correspond to the results of $\Delta_1=0.5$.  \label{fanoline}}
\end{figure}

\begin{figure}[htb]
\centering \scalebox{0.16}{\includegraphics{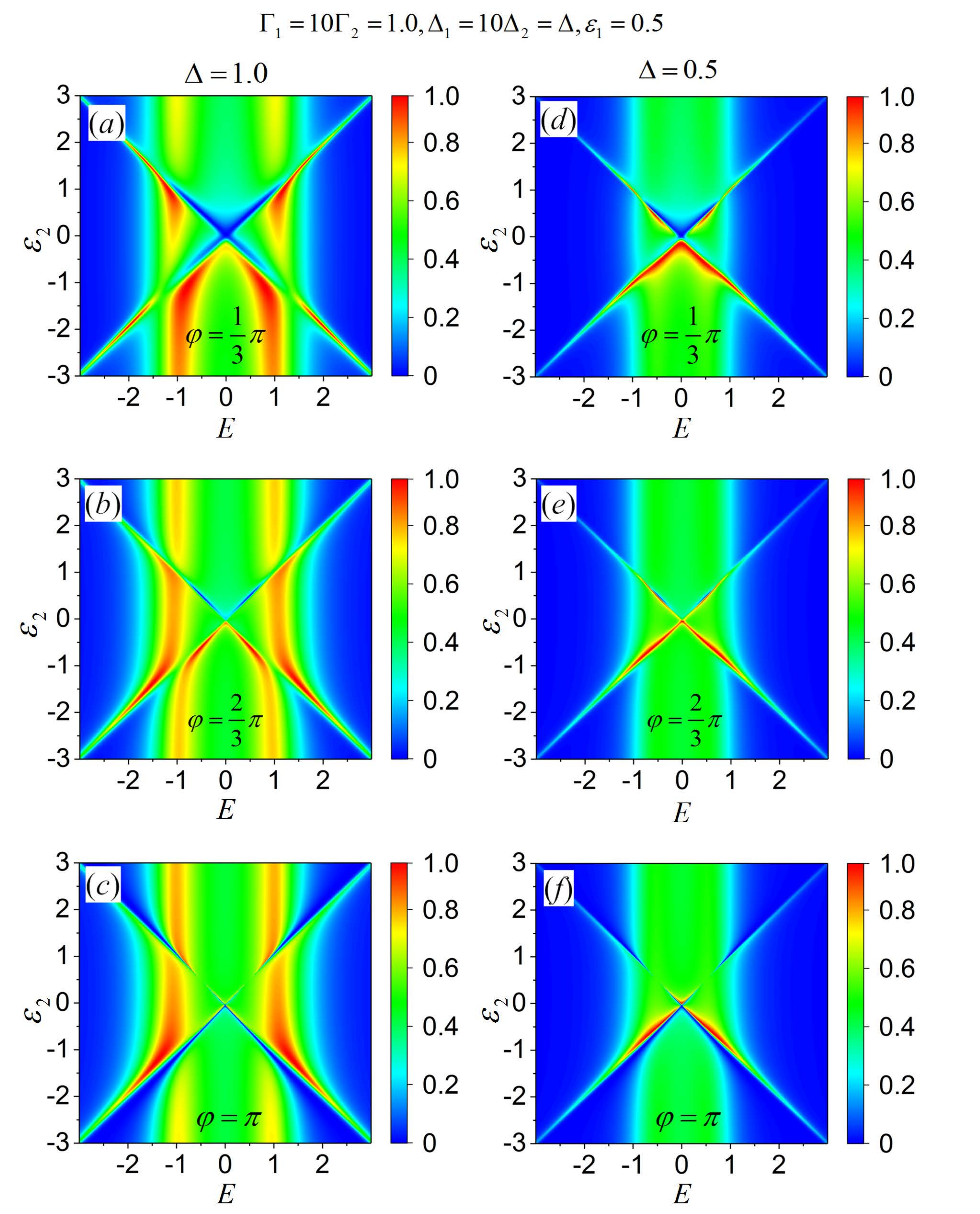}} \caption{
AR spectra affected by the local magnetic flux through this system. $\varepsilon_1$ is assumed to be 0.5. The other parameters are identical with those in Fig.7, respectively.  \label{fanoline}}
\end{figure}

\par
In this part, we would like to pay attention to the case of asymmetric QD-lead couplings, to investigate the AR behaviors. One reason is that the quantum transport properties in the parallel double-QD structure is tightly dependent on the QD-lead coupling manners. Without loss of generality, the QD-lead(SC) couplings are taken to be ${\Gamma_1\over\Gamma_2}={\Delta_1\over\Delta_2}=\lambda=10$. In this case,
\begin{equation}
\tau(E)=\frac{(1+\lambda-\sqrt{\lambda})E^2-(\lambda\varepsilon_1-\varepsilon_2)^2}{\det|[G^r(E)]^{-1}|}\Gamma_2\Delta_2 \notag
\end{equation}
in the absence of magnetic flux. As a result, the antiresonance positions shift to the points of $E=\pm{\lambda\varepsilon_1-\varepsilon_2\over\sqrt{1+\lambda-\sqrt{\lambda}}}$.
\par
Fig.6 shows the AR spectra in the cases of $\Delta_1=1.0$ and $0.5$, respectively, by taking the magnetic flux out of account.
We can see that the asymmetry of the QD-lead couplings takes nontrivial effect to the AR process of this system. Firstly, in Fig.6(a) it shows that with the change of $E$, two main peaks appear in the AR spectrum, near the regions of $E=\pm1.0$. The position of the AR peaks seem to be independent of the shift of $\varepsilon_2$. The other phenomenon is that the AR spectrum is clearly divided into four regions by the two antiresonance conditions shown in the above paragraph. Accordingly, the value of $T_A(E)$ is exactly equal to zero at the energy zero point. With the increase of $\varepsilon_1$, the asymmetric AR spectrum comes into being with the shift of $\varepsilon_2$, as shown in Fig.6(b)-(c). Meanwhile, the edges of the four AR regions becomes unclear gradually, though the lines can not be obviously modified by the change of $\varepsilon_2$. Next, the case of $\Delta_1=0.5$ is exhibited in Fig.6(d)-(f). In the case of $\varepsilon_1=0$, one can find one wide peak around the point of $E=0$. With the increase of $\varepsilon_1$, this peak is destroyed, and the suppression of the AR process can also be observed, accompanied by the appearance of the asymmetric AR spectrum. Therefore, the asymmetry of the QD-lead coupling can efficiently alter the AR properties in the considered structure.

\par
In Fig.7, we continue to present the influence of the local magnetic flux through this ring on the AR properties. Fig.7(a)-(c) show the results of $\varphi={\pi\over3}$, $2\pi\over3$, and $\pi$, respectively, in the case of $\varepsilon_1=0.5$ with $\Delta_1=1.0$. And Fig.4(d)-(f) correspond to the result of $\Delta_1=0.5$. In this figure, one can find that with the increase of magnetic flux, the AR spectrum exhibits obvious changes. Similar to the results in Fig.4, the $E_1$-contributed AR peaks are strengthened and widened apparently. In addition, the AR maximum near the positions of $E=\pm1.0$ is suppressed, compared with the case of $\varphi=0$. The other result is that due to the disappearance of the antiresonances, the partition of the AR spectrum becomes ambiguous in the cases of $\varphi={\pi\over3}$ and ${2\pi\over 3}$. Next in Fig.7(d)-(f), it shows that the magnetic flux indeed reverses the symmetry of the AR spectrum, but the AR ability cannot be enhanced in this process.

\begin{figure}
\centering \scalebox{0.41}{\includegraphics{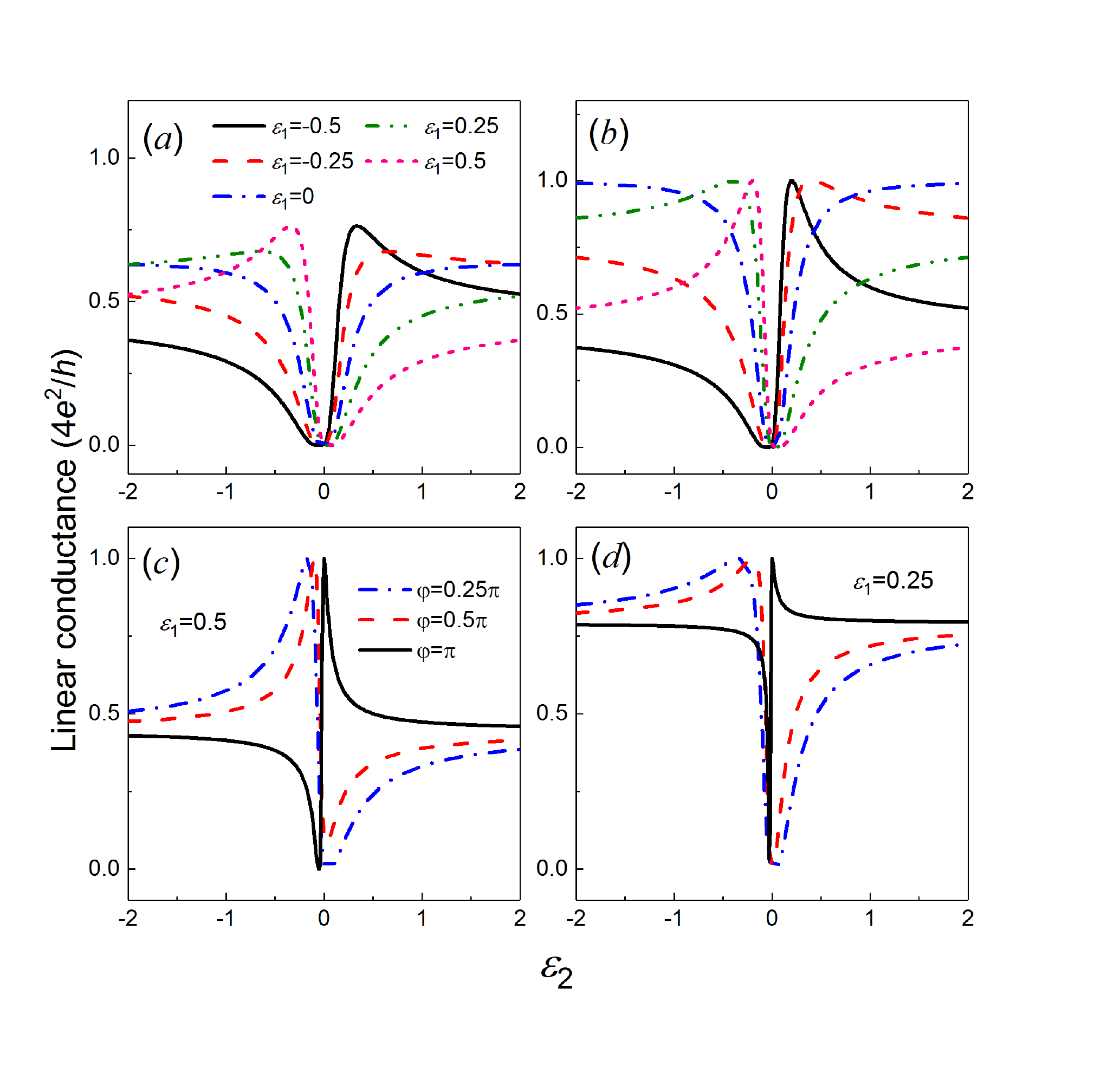}} \caption{
Linear-conductance spectra of the AR in the parallel double-QD structure. (a)-(b) Results in the cases of $\Delta_1=1.0$ and $0.5$. The level of QD-1 is taken to be $\varepsilon_1=\pm0.5$, $\pm0.25$, and $0$, respectively. (c)-(d) Impact of magnetic flux on the conductance spectra. \label{fanoline}}
\end{figure}

\par
The results in Fig.6 and Fig.7 show that the antiresonance is allowed to take place even in the case of $E=0$ when $\varphi=n\pi$. Thus we anticipate that the Fano effect can be observed in the linear conductance spectrum despite the asymmetry of the QD-lead couplings. With this idea, we plot the linear conductance profile of the AR in Fig.8. It is clearly shown that apparent Fano lineshape exists in the conductance spectrum. As shown in Fig.8(a)-(b), the Fano effect is enhanced when the superconducting pairing potentials in the QDs decrease to $\Delta=0.5$. This is due to the fact that in this case, the Fano resonance is able to to take place. Also, compared with the result in Fig.5, the reference channel makes more contribution to the Fano effect in this case, because of the increase of the conductance value outside the Fano interference region. In addition, in Fig.8(c)-(d), it shows that the local magnetic flux can reverse the Fano lineshape in the conductance spectrum. Up to now, we have known that in this structure, the Fano effect is robust in the linear AR process when one arm provides the reference channel.

\subsection{Analysis on the Fano effect}
\par
We next aim to clarify the Fano effect in the linear AR process. It is well-known that for discussing the Fano effect, one has to transform the conductance expression into its Fano form. Following this idea, we start to analyze the Fano effect in the linear AR process by presenting the analytical expression of the AR coefficient. After a series of deduction, we are allowed to obtain the Fano expression of $\tau|_{E=0}$, i.e.,
$\tau|_{E=0}=\tau_b\frac{\alpha(e+q)^2}{\alpha e^2+1}e^{i\varphi}$.
And then,
\begin{equation}
T_A|_{E=0}=|\tau_b|^2\frac{\alpha^2|e+q|^4}{|\alpha e^2+1|^2}.
\end{equation}
Here $\tau_b=\frac{4\Gamma_1\Delta_1}{4\varepsilon_1^2+\Gamma_1^2+4\Delta_1^2 }$ is the AR coefficient in the absence of QD-2. $e=\varepsilon_2+\tau_b\frac{\Gamma_1\Gamma_2+4\Delta_1\Delta_2}{4\Gamma_1\Delta_1}\varepsilon_1$, and
\begin{equation}
q=-(\frac{\Gamma_1\Gamma_2+4\Delta_1\Delta_2}{4\Gamma_1\Delta_1}\tau_b+\frac{\sqrt{\Gamma_2\Delta_2}} {\sqrt{\Gamma_1\Delta_1}}e^{-i\varphi})\varepsilon_1
\end{equation}
is the so-called Fano parameter. In addition, the other quantity also comes into play, i.e., $\alpha=\frac{16\Gamma_1^2\Delta_1^2}{[4(\Gamma_1\Delta_2-\Delta_1\Gamma_2)^2+4\varepsilon_1^2(\Gamma_2^2+4\Delta_2^2)]\tau_b^2+4\Gamma_1\Delta_1\tau_b\cal C}$, which is always positive. It should be noticed that this Fano expression is basically complicated and different from the result in the case of single electron tunneling where $
T_{et}=T_b\frac{|e+q|^2}{e^2+1}$ with $T_b$ being the electron tunneling ability in the reference channel \cite{Fano1,Fano2,Fano3}. This reflects the complicated nature of the Fano effect in linear AR process.
\par

In Eq.(16), one can find that in this structure, the arm of QD-1 enables to provide the reference channel for the Fano interference in the linear AR process. However, the information of the resonant channel cannot be directly obtained from the other arm, due to the complexity of the parameter $e$. Surely, with the change of $e$, the profile of $T_A|_{E=0}$ indeed exhibits the Fano lineshape, for a real Fano parameter $q$. And when the sign of $q$ (i.e., $\pm$) changes, the Fano lineshape will be reversed. One can then understand the Fano lineshapes in the linear conductance spectra in Fig.5 and Fig.8.
The Fano antiresonance can be analyzed as follows. When finite magnetic flux is introduced through this ring, the Fano parameter $q$ will become complex, so $e+q\neq0$ and the Fano effect will be suppressed. Otherwise, in the cases of $\varphi=0(\pi)$, the position of the Fano antiresonance can be ascertained by letting $e+q=0$, and the corresponding condition can be solved, i.e.,
\begin{equation}
\varepsilon_1\sqrt{\Gamma_2\Delta_2}\mp\varepsilon_2\sqrt{\Gamma_1\Delta_1}=0.
\end{equation}
This relationship is simplified to be $\varepsilon_1\mp\varepsilon_2=0$ in the case of $\Delta_j=\Delta$ with the same QD-lead couplings. According to these results, we know that for a fixed $\varepsilon_1$, there always exists the antiresonance point in the linear conductance spectrum, with its position being at the points of $\varepsilon_2=\pm\sqrt{\frac{\Gamma_2\Delta_2}{\Gamma_1\Delta_1}}\varepsilon_1=\pm\lambda\varepsilon_1$ in the case of $\varphi=0(\pi)$.
\begin{figure}
\centering \scalebox{0.38}{\includegraphics{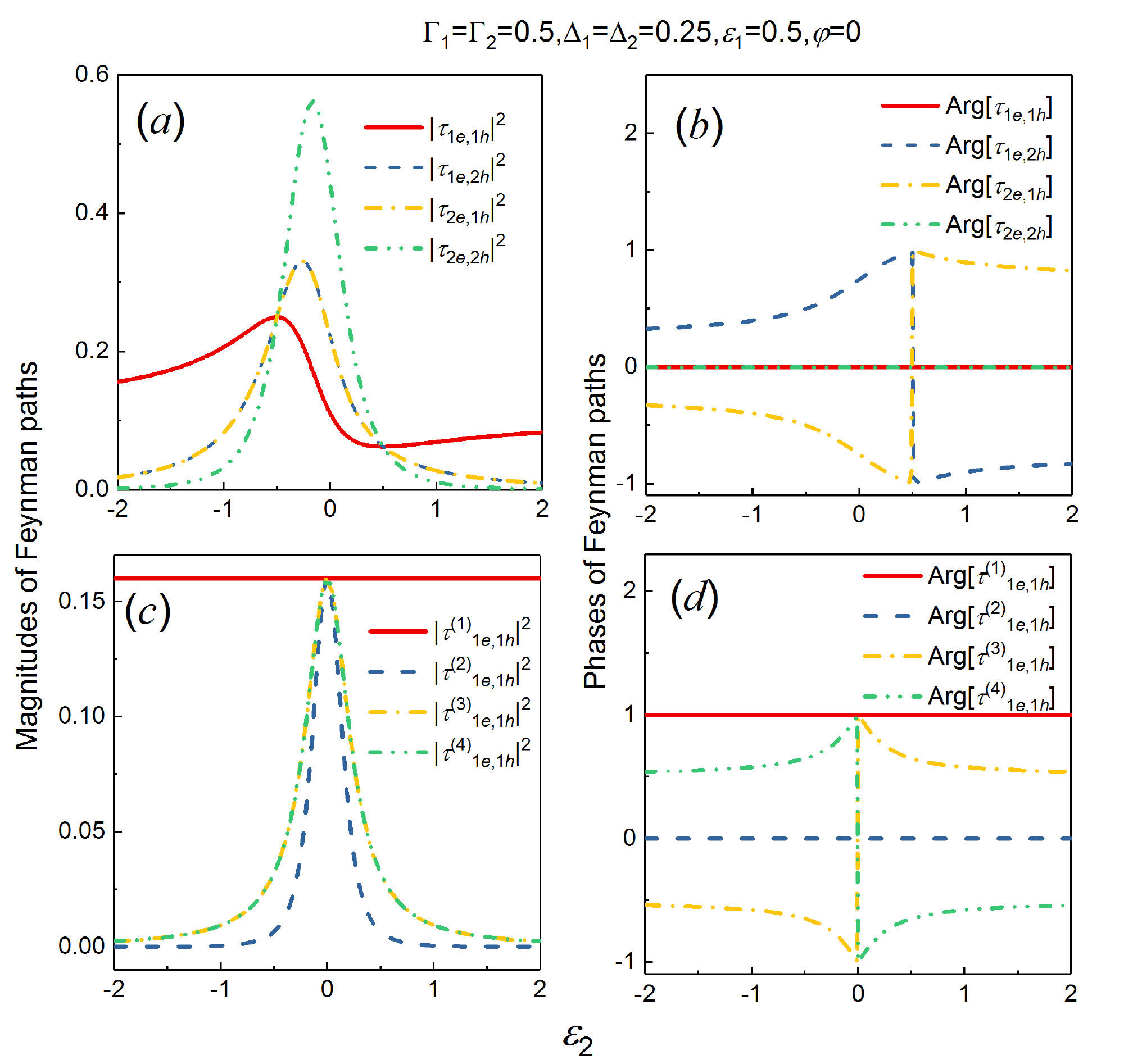}} \caption{
(a)-(b) Magnitudes and phases of the Feynman paths $\tau_{je,lh}$. (c)-(d) Magnitudes and phases of the zero-order paths of $\tau^{(j)}_{1e,1h}$. Related parameters are taken to $\Gamma_j=0.5$, $\Delta_j=0.25$, and $\varepsilon_1=0.5$. \label{fanoline}}
\end{figure}

\par
It should be admitted that the Fano effect in the linear AR process stems from the quantum interference among the Feynman paths involved. Thus we need to clarify the interference manner to further understand the Fano effect in this system. To begin with, it is not difficult to find that $\tau=\tau_{1e,1h}+\tau_{2e,1h}+\tau_{1e,2h}+\tau_{2e,2h}$
where $\tau_{je,lh}={\cal V}_{je}G^r_{je,lh}{\cal V}_{lh}$ with ${\cal V}_{1e(h)}=\sqrt{\Gamma_1}e^{\mp i\varphi/2}$, and ${\cal V}_{2e(h)}=\sqrt{\Gamma_2}e^{\pm i\varphi/2}$. This means that the Fano interference takes place among these four paths. Their contributions to the linear conductance are shown in Fig.9(a). One can see the nonzero contribution of each path. In addition to $\tau_{2e,2h}$, the other two paths, i.e., $\tau_{2e,1h}$ and $\tau_{1e,2h}$, also exhibit Breit-Wigner lineshapes with the change of $\varepsilon_2$, despite their different magnitudes. One can then know that the resonant channel of the Fano interference is indeed offered by the coupling between QD-2 and the normal lead and SC, whereas $\tau_{2e,1h}$ and $\tau_{1e,2h}$ provide the supplemental but necessary information on this channel. The results in Fig.9(b) describe the phases of the four paths. They can help us to understand the Fano resonance and antiresonance points in the linear conductance spectrum. On the other hand, in Fig.9(a) we have to notice that $\tau_{1e,1h}$ does not only provide the reference channel, due to its asymmetric lineshape in the interference region. Therefore, it is necessary to present its detailed content by writing it as $\tau_{1e,1h}=\sum_{j=1}^4\tau^{(j)}_{1e,1h}$ with
\begin{small}
\begin{eqnarray}
&&\tau^{(1)}_{1e,1h}= {\cal V}_{1e}g_{1e}\Delta_1g_{1h} \tilde{A}_p{\cal V}_{1h};\notag\\
&&\tau^{(2)}_{1e,1h}=-{1\over 4 }{\cal V}_{1e}g_{1e}{\cal V}^*_{1e}{\cal V}_{2e}g_{2e}\Delta_2g_{2h}{\cal V}_{2h}{\cal V}^*_{1h}g_{1h} \tilde{A}_p{\cal V}_{1h};\notag\\
&&\tau^{(3)}_{1e,1h}={i\over 2 }{\cal V}_{1e}g_{1e}\sqrt{\Delta_1\Delta_2}g_{2h}{\cal V}_{2h}{\cal V}^*_{1h}g_{1h} \tilde{A}_p{\cal V}_{1h};\notag\\
&&\tau^{(4)}_{1e,1h}={i\over 2 }{\cal V}_{1e}g_{1e}{\cal V}^*_{1e}{\cal V}_{2e}g_{2e}\sqrt{\Delta_1\Delta_2}g_{1h} \tilde{A}_p{\cal V}_{1h}.\notag
\end{eqnarray}
\end{small}
Here $ \tilde{A}_p=1/[g_{1e}g_{1h}g_{2e}g_{2h} \det|[G^r]^{-1}|_{E=0}]=\sum_{l=0}^{\infty} A^{(l)}=1-{i\over 2}\sqrt{\Delta_1\Delta_2}g_{2e}\Delta_2g_{2h}{\cal V}_{2h}{\cal V}^*_{1h}g_{1h}+\cdots$. One can then understand the complicated quantum interference manner, by recognizing the multiple Feynman paths. And it is certain that the interference among them leads to the asymmetric lineshape of $|\tau_{1e,1h}|^2$. Via a further observation, it can be known that $\tau^{(1)}_{1e,1h}$ makes a leading contribution to the interference, since its zero-order component exactly describes the particle motion in the reference channel. Therefore, the arm of QD-1 provides the reference channel but not only the reference channel for the Fano interference of our considered system. In Fig.9(c)-(d), the magnitudes and phases of the zero-order paths of $\tau^{(j)}_{1e,1h}$ are presented. With the help of such analysis, we can understand that
the direct AR processes in the two arms provide the reference and resonant channels for the Fano interference. It is the complicated interference among multiple Feynman paths that drives the appearance of the Fano lineshape in the linear conductance spectrum.
\par
\begin{figure}[htb]
\centering \scalebox{0.39}{\includegraphics{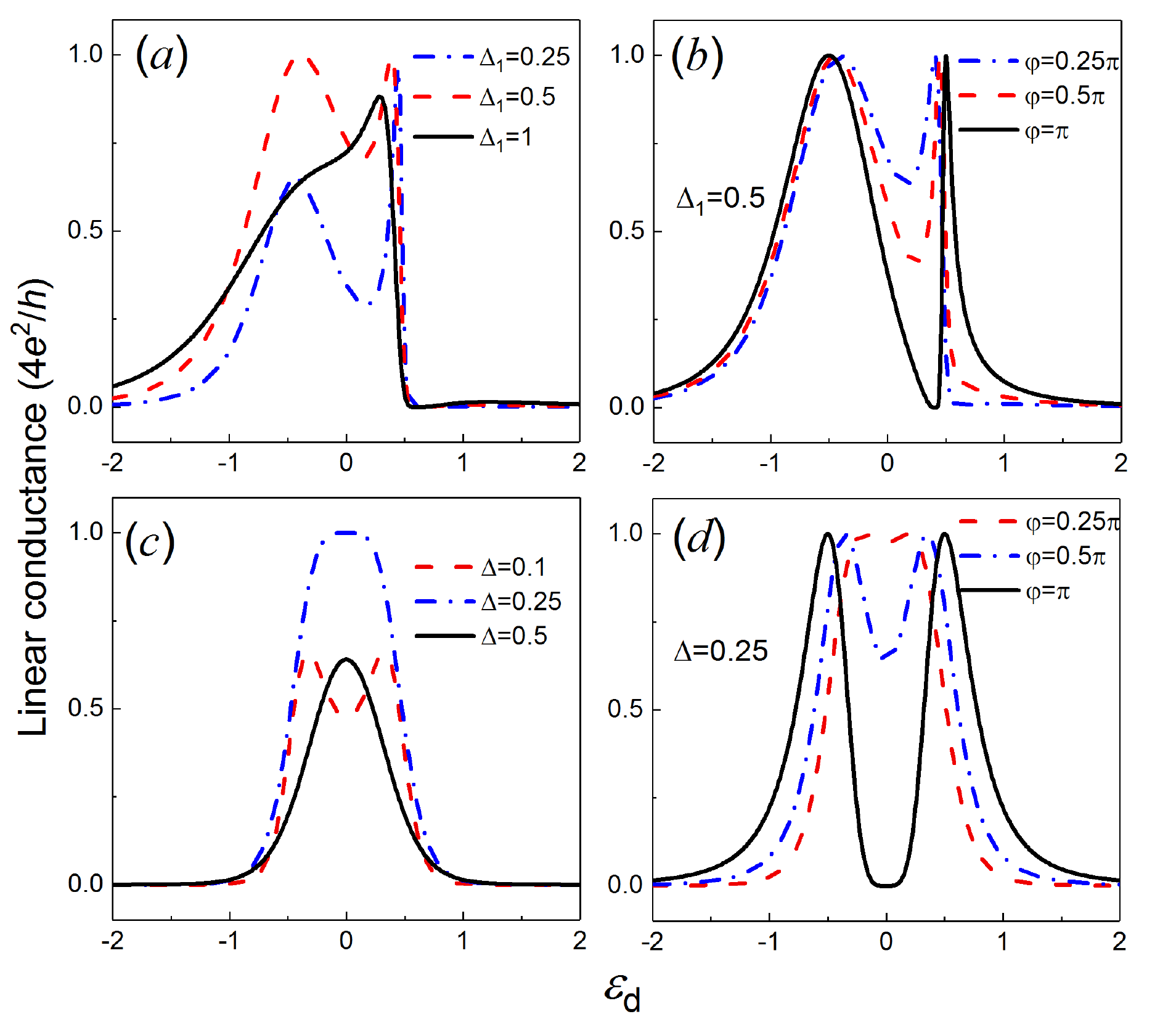}} \caption{
Linear-conductance spectra of the Andreev reflection in the case of $\varepsilon_j=\varepsilon_d\pm\delta$ with $\delta=0.5$. (a)-(b) Results of $\Gamma_1=10\Gamma_2$ and $\Delta_1=10\Delta_2$. (c)-(d) The QD-lead(SC) couplings are taken to be the same, i.e., $\Gamma_j=\Gamma$ and $\Delta_j=\Delta$. \label{fanoline}}
\end{figure}
\par

\par
Since the two arms can provide the reference and resonant channels for the Fano interference, we would like to investigate the Fano effect in the case where the levels of the two QDs are tunable, e.g., $\varepsilon_1=\varepsilon_d+\delta$ and $\varepsilon_2=\varepsilon_d-\delta$ by taking $\Gamma_1=10\Gamma_2$ and $\Delta_1=10\Delta_2$. One expects that due to the asymmetric QD-lead coupling, the Fano effect can be driven in this case.
As shown in Fig.10(a), in the case of $\Delta_1=0.25$, two peaks appear in the linear-conductance spectrum. One is ``more" resonant and the other is ``less" resonant. This is exactly the necessary condition of the Fano interference. Thus, one sees the occurrence of the Fano effect. When $\Delta_1=0.5$, the conductance magnitude increases, and the Fano effect becomes apparent. However, the further increase of $\Delta_1$ can only weaken the Fano effect [See the case of $\Delta=1.0$]. Next, after the application of the magnetic flux, the Fano lineshape in the conductance spectrum can be reversed in the case of $\varphi=\pi$ [See Fig.10(b)]. One can then find that even when the two QDs are both level-tunable, the Fano effect can be induced in the linear AR process, in the case of the asymmetric QD-lead couplings.
\par
For comparison, in Fig.10(c)-(d) we consider the other case, i.e., $\Gamma_j=\Gamma$ and $\Delta_j=\Delta$, by considering $\varepsilon_1=\varepsilon_d+\delta$ and $\varepsilon_2=\varepsilon_d-\delta$ with $\delta$. As shown in Fig.10(c), in the case of $\Delta=0.1$, two peaks appear in the linear-conductance spectrum. With the increase of $\Delta$ to $\Delta=0.25$, one peak arises at the position of $\varepsilon_d=0$ in the conductance spectrum. When $\Delta$ is further increased to $\Delta=0.5$, the conductance peak is weakened. Thus in such a case, no Fano effect takes place. The underlying physics should be attributed to the disappearance of the reference channel. When the magnetic flux is taken into account, the conductance spectrum can be changed, as shown in Fig.10(d). In the case of $\varphi=\pi$, one antiresonance point can be observed at the position of $\varepsilon_d=0$, but it cannot be viewed as the Fano antiresonance.

\section{summary\label{summary}}
To sum up, in this work we have investigated systematically the AR properties in the parallel double-QD structure, by considering one metallic lead and one $s$-wave SC to couple to the QDs respectively. As a result, it has been found that Fano lineshhapes are allowed to appear in the linear conductance spectrum of the AR process, when one arm provides the reference channel for the quantum interference. Besides, the Fano lineshapes can be reversed by tuning the QD level or local magnetic flux. However, the condition of the Fano resonance is completely different from that in the normal electron transport case. By presenting the Fano form of the linear-conductance expression, we have presented a detailed explanation about the Fano effect in this structure. And the Fano interference manner has been described by means of the Feynman path language. We believe that this work can be helpful for understanding the Fano interference in the AR process.

\section*{Acknowledgements}
This work was financially supported by the Liaoning BaiQianWan Talents Program (Grant No. 201892126) and the Fundamental Research
Funds for the Central Universities (Grant No. N160504009 and N170506007).

\clearpage

\bigskip

\end{document}